\documentclass{article}

\usepackage{epsfig}
\usepackage{amssymb,latexsym}

\newcommand{\beq}{\begin{equation}}
\newcommand{\eeq}{\end{equation}}
\newcommand{\bqa}{\begin{eqnarray}}
\newcommand{\eqa}{\end{eqnarray}}
\newcommand{\fr}{\frac}
\renewcommand{\baselinestretch}{1.67}

\begin{document}
\title{\bf \Large Singularities in gravitational collapse with radial pressure}
\author{{\large S\' ergio M. C. V. Gon\c calves\footnote{E-mail: sergiog@its.caltech.edu, Tel: +1-626-395-8753, Fax: +1-626-796-5675}} \\
{\em \normalsize Theoretical Astrophysics, California Institute of Technology,
Pasadena, California 91125} \\
\\
{\large Sanjay Jhingan} \\
{\em \normalsize Yukawa Institute for Theoretical Physics, Kyoto University,
Kyoto 606-8502, Japan}}
\date{\small \today}

\maketitle
\begin{abstract}
We analyze spherical dust collapse
with non-vanishing radial pressure, $\Pi$, and vanishing tangential
stresses. Considering a barotropic equation of state, $\Pi=\gamma\rho$, we obtain an analytical solution in closed form---which is exact for $\gamma=-1,0$, and approximate otherwise---near the center of symmetry (where the curvature singularity forms). We study the formation, visibility, and curvature strength of singularities in the resulting spacetime. We find that visible, Tipler strong singularities can develop from generic initial
data. Radial pressure alters the spectrum of possible endstates for
collapse, increasing the parameter space region that contains no
visible singularities, but {\em cannot} by itself prevent the formation of
visible singularities for sufficiently low values of the energy
density. Known results from pressureless dust are recovered in the $\gamma=0$ limit.\\
\\
Keywords: Gravitational collapse, singularities, black holes.
\end{abstract}
\newpage
\section{Introduction}
\renewcommand{\baselinestretch}{1.67}
{\small $\;$}

It has long been known that under a variety of
circumstances, spacetimes which are solutions of Einstein's equations
with physically reasonable regular initial data, inevitably develop
singularities \cite{hawking&ellis73}. These are events at which
Riemannian curvature typically diverges, the spacetime is geodesically
incomplete, and classical general relativity necessarily breaks
down. The question of whether these singularities can be visible is
one of the outstanding problems in general relativity.

In an effort to protect the applicability of general relativity,
Penrose proposed that such singularities might be hidden by an event
horizon, and thus invisible to an asymptotic observer, i.e., they
cannot be {\em globally} naked \cite{penrose69}. This constitutes in
essence what has become known as the {\em weak cosmic censorship
conjecture.} However, it is quite possible---at least in
principle---for an observer to penetrate the event horizon and live a
rather normal life inside a black hole. This motivated the
{\em strong cosmic censorship conjecture}, which broadly states that
timelike singularities cannot occur in nature, i.e., they cannot be visible even {\em locally} \cite{penrose79}.

A lack of tools to handle global properties of the Einstein equations
(and respective solutions), together with their high non-linearity,
have been the main obstacle to provable formulations of either form of the cosmic censorship conjecture. Whilst efforts are being
undertaken in this direction \cite{global}, one can hope that the
detailed study of specific models helps to isolate some defining
features of singularity formation and structure, thereby contributing
towards a precise, counter-example-free formulation of the
conjecture.

One such model is Lema\^{\i}tre-Tolman-Bondi (LTB) \cite{ltb}
inhomogeneous dust collapse, whose general solution is analytically obtainable
in closed (or parametric) form. The exact solvability of this model
has resulted in many detailed studies by various authors
\cite{tbauthors,tbauthors2}. These analyses show that, from generic
initial data, a
null (central) singularity develops, which is Tipler strong
\cite{tipler77}, and can be locally or globally naked, depending on
the differentiability of the initial density profile at the center
\cite{jhingan}. In
addition, the singularity is stable against initial density
perturbations \cite{deshingkar&joshi&dwivedi99}, and marginally stable
against linear non-spherical perturbations \cite{hin98-00}. It is fair
to say that singularity formation and structure in LTB collapse are
now very well understood.

Whilst spherical symmetry may, arguably, constitute a reasonable approximation
to realistic collapse \cite{nakamura&sato82}, pressureless dust
evidently fails to be a realistic form of matter, especially at the
late stages of stellar collapse, where an effective equation of state
must be considered, and radial and tangential stresses come into play
\cite{miller&sciama80}. The inclusion of pressure has been studied
analytically in the perfect fluid and anisotropic pressure cases; the
former, when self-similarity \cite{joshi&dwivedi92-93,carr&coley00},
or special equations of state \cite{rocha&wang&santos99} are assumed.

Early analytical studies of perfect fluid collapse were
restricted to approximate solutions of Einstein's equations near the
singularity
\cite{podurets}. This technique was generalized to include a barotropic
equation of state, $p=k\rho$, for perfect fluids, but the
results were restricted to the analysis of non-central shells due to
simplifying assumptions \cite{cooperstock}. Numerical studies of
self-similar \cite{ori&piran87&90} and perfect fluid collapse with
barotropic equation of state \cite{harada98} have shown the formation
of a central visible (globally naked, in the self-similar case)
singularity. The work of Joshi and Dwivedi \cite{cmp94} showed the
possibility of naked singularity formation in the collapse of general ``type
I'' matter fields \cite{matter}, but their analysis does not provide an
explicit relation between the initial data---such as densities and
pressures---and the conditions for formation of visible singularities; 
(such a relation is very clear in the LTB case).

Recently, models with only tangential stresses have been analyzed. In
particular, the Einstein cluster class is now completely understood
\cite{Magli}. Tangential stresses tend to uncover part of the
singularity spectrum, thereby showing that the role of tangential
stresses is not negligible, and lending support to the idea that naked
singularities develop from generic gravitational collapse
\cite{MagliII}.

In this paper, we present an new solution of spherical dust
collapse with non-vanishing radial pressure, and vanishing tangential
stresses. This solution is exact for $\gamma=-1$ and $\gamma=0$. For other values of $\gamma\in(-1,1)$, it asymptotes the exact solution in the $r=0$ limit, which is the region of interest, where the curvature singularity forms. In particular, the solution can be made to obey Einstein's equations to arbitrarily small error, by considering an arbitrarily small neighborhood of the $r=0$. In this way, the role of radial pressure---as opposed to the
combined contribution of radial and tangential stresses, as in the
perfect fluid case---is isolated and its effects on singularity
formation and structure become clear. For configurations with radial
pressure $\Pi$ linearly proportional to the density ($\Pi = \gamma \rho$),
that can end
up in either a black hole or naked singularity, an increase in radial
pressure leads to a decrease in the parameter space area
for naked singularities. In the initial two-parameter data space (radial pressure vs. density),
visible singularities are shown to exist for the entire range of the
adiabatic index $\gamma$. This shows that, whereas radial pressure can
cover singularities that would otherwise be visible, it cannot, by
itself prevent the formation of visible singularities.

This paper is organized as follows. In Sec. II, Einstein's equations
are solved for a spherically symmetric system with non-vanishing radial
pressure. The ansatz of marginally bound configurations yields a
complete analytical treatment in closed form. In Sec. III, conditions
for singularity formation are discussed. Section IV studies the
visibility of the central curvature singularity and discusses the role
of radial pressure. Curvature strength is computed in Sec. V. Section
VI discusses the pure dust limit ($\gamma=0$) of our model. Section
VII concludes with a summary and discussion.

Geometrized units, in which $G=c=1$, are used throughout.

\section{Spherically symmetric collapse with radial pressure}

We consider a general spherically symmetric metric, in standard
spherical coordinates $\{t,r,\theta,\phi\}$:
\beq
ds^{2}=-e^{2\Phi(t,r)}dt^{2}+e^{-2\Psi(t,r)}dr^{2}+R^{2}(t,r)d\Omega^{2}, \label{met}
\eeq
where $R(t,r)$ is the proper area radius and
$d\Omega^{2}=d\theta^{2}+\sin^{2}\theta d\phi^{2}$ is the
canonical metric of the unit two-sphere.

The stress-energy tensor is:
\beq
T_{a}^{b}=\mbox{diag}(-\rho,\Pi,\Sigma,\Sigma)
\eeq
where $\rho$, $\Pi$, and $\Sigma$ are the energy density, radial
pressure, and tangential stress, respectively.

With the metric (\ref{met}) the non-vanishing Einstein tensor
components are:
\bqa
G_{tt}&=&\fr{e^{2(\Psi+\Phi)}}{R^{2}}\left[2R(R''+R'\Psi')+R'^{2}-e^{-2\Psi}\right]-\fr{\dot{R}}{R}\dot{\Psi}+\left(\fr{\dot{R}}{R}\right)^{2}, \\
G_{tr}&=&\fr{2}{R}\left(\dot{R}'-\dot{R}\Phi'+\dot{\Psi}R'\right), \\
G_{rr}&=&\fr{e^{-2(\Psi+\Phi)}}{R^{2}}\left[2R(\dot{R}\dot{\Phi}-\ddot{R})-\dot{R}^{2}-e^{2\Phi}\right]+\fr{R'}{R^{2}}(R'+2R\Phi'), \\
G_{\theta\theta}&=&-R^{2}
\left\{e^{-2\Phi}\left[-\fr{\dot{R}}{R}(\dot{\Phi}+\dot{\Psi})+\fr{\ddot{R}}{R}+\dot{\Psi}^{2}-\ddot{\Psi}+\dot{\Phi}\dot{\Psi}\right]\right. \nonumber \\
&&\left.-e^{2\Psi}\left[\fr{R'}{R}(\Phi'+\Psi')+\fr{R''}{R}+\Phi'^{2}+\Phi''+\Phi'\Psi'\right]\right\},
\\
G_{\phi\phi}&=&\sin^{2}\theta G_{\theta\theta},
\eqa
where $'\equiv\partial_{r}$ and $\dot{}\equiv\partial_{t}$.

Introducing the auxiliary functions
\bqa
k(t,r)&\equiv&1-e^{2\Psi}R'^{2}, \label{kay} \\
m(t,r)&\equiv&\fr{1}{2}R\left(e^{-2\Phi}\dot{R}^{2}+k\right), \label{mas}
\eqa
Einstein's equations can be recast as
\bqa
m'&=&4\pi R^{2}R'\rho, \label{mpr} \\
\dot{m}&=&-4\pi R^{2}\dot{R}\Pi, \label{mdt} \\
\Phi'&=&(\rho+\Pi)^{-1}\left[2\fr{R'}{R}(\Sigma-\Pi)-\Pi'\right], \label{Phpr} \\
\dot{\Psi}&=&\fr{\dot{R}}{R'}\Phi'-\fr{\dot{R}'}{R'}. \label{Pspr}
\eqa
Note that the function $m(t,r)$ is just the Misner-Sharp mass
\cite{misner&sharp64}:
\beq
m(t,r)=\fr{R}{2}\left(1-R_{,a}R_{,b}g^{ab}\right).
\eeq
It then follows that the function $k(t,r)$ is the binding energy  per
unit mass of a shell $r$, with area radius $R(t,r)$. Gravitationally
bound configurations have $0<k<1$; unbound configurations have $k<0$,
and $k=0$ corresponds to the marginally bound case.

From Eq. (\ref{mas}), an ``acceleration'' equation for the area radius
can be easily obtained:
\beq
\ddot{R}=\dot{\Phi}\dot{R}-e^{2\Phi}\left[4\pi\Pi
R+\fr{m}{R^{2}}+(k-1)\fr{\Phi'}{R'}\right]. \label{ae}
\eeq

When the tangential stresses vanish, $\Sigma=0$, Eq. (\ref{Phpr})
simplifies to:
\beq
\Phi'=-(\rho+\Pi)^{-1}\left(2\fr{R'}{R}\Pi+\Pi'\right). \label{Phpr2}
\eeq

\subsection{Marginally bound models}

Let us consider the case $k=0$, corresponding to marginally bound
configurations. From Eq. (\ref{kay}), provided $R'>0$ (i.e.,
absence of shell-crossings), it follows that
\beq
\dot{\Psi}=-\fr{\dot{R}'}{R'},
\eeq
which implies, from Eq. (\ref{Pspr}),
\beq
\Phi'=0, \label{phipz}
\eeq
i.e., $\Phi=\Phi(t)$.

We obtain a closed system by specifying an equation of state, which we
take to be of barotropic form:
\beq
\Pi=\gamma\rho, \label{estate}
\eeq
where causality requires $\gamma^{2}<1$. From Eqs. (\ref{Phpr2}) and
(\ref{phipz}) it follows that
\beq
\rho(t,r)=\fr{\omega(t)}{R^{2}(t,r)}, \label{rp}
\eeq
where $\omega(t)$ is an arbitrary real-valued function of $r$.

Equations
(\ref{mpr})-(\ref{mdt}) now read
\bqa
m'&=&4\pi\omega R', \label{mpr0} \\
\dot{m}&=&-4\pi\omega\gamma \dot{R}. \label{mdt0}
\eqa
Equation (\ref{mpr0}) readily integrates to
\beq
m(t,r)=4\pi\omega R, \label{msm}
\eeq
where the ``constant'' of integration (a function of $t$ alone) was set to zero by demanding
regularity at the
center. Taking the partial time derivative of the above equation and
equating it to
Eq. (\ref{mdt0}) gives
\beq
\dot{\omega}R + (1+\gamma)\omega\dot{R}=0.
\eeq
Which integrates to
\beq
\omega(t)=\omega_{0}(r)R^{-1-\gamma},
\eeq
where $\omega_{0}(r)$ is an integration ``constant'', which can be
fixed as $\omega_{0}(r)=Cr^{1+\gamma}$, by the choice $R(0,r)=r$, 
where $C\geq0$. Hence,
\beq
\omega(t)=C\left(\fr{r}{R}\right)^{1+\gamma}.
\eeq
From Eq. (\ref{estate}) we then have
\beq
\Pi(t,r)=\fr{\gamma C}{R^{2}}\left(\fr{r}{R}\right)^{1+\gamma}.
\eeq
Note that $C$ characterizes the ``strength'' of the initial density
profile---which is, by construction, positive definite, thus enforcing
the weak energy condition---whereas $\gamma\in(-1,1)$ characterizes
the relative strength of the radial pressure (which can be
negative). If we regard $\Pi$ as a hydrostatic pressure, then
$\sqrt{\gamma}$ is the sound speed, which is complex for negative
pressures. Since we do not know how matter behaves in high density
regimes, such as the strong field regions surrouding the central
singularity, we allow $\gamma$ to be negative, within the limits
imposed by the weak energy condition:
\beq
\rho+\Pi\geq0\Rightarrow\gamma\geq-1.
\eeq
(Note that causality implies that this inequality does not saturate).

The mass function becomes
\beq
m(t,r)=4\pi C R\left(\fr{r}{R}\right)^{1+\gamma}, \label{mass}
\eeq
and the evolution equation simplifies to
\beq
e^{-2\Phi}\dot{R}^{2}=8\pi C \left(\fr{r}{R}\right)^{1+\gamma}. \label{eveq}
\eeq
Now, since $\Phi=\Phi(t)$, one can trivially rescale the comoving time
$t$ to proper time $\tau$, via
\beq
d\tau=e^{\Phi(t)}dt\;\Rightarrow\;\tau=\int e^{\Phi(t)}dt+f(r).
\eeq
Setting $f(r)=0$, Eq. (\ref{eveq}) becomes
\beq
\dot{R}^{2}(\tau,r)=8\pi
C \left(\fr{r}{R}\right)^{1+\gamma}, \label{Rdt}
\eeq
where the dot denotes partial differentiation with respect to $\tau$,
and $R$ is to be regarded as a function of the independent coordinates $\tau$ and $r$. Equation
(\ref{Rdt}) integrates to
\beq
R(\tau,r)=A^{\fr{1}{3+\gamma}}r^{\fr{1+\gamma}{3+\gamma}}
\left[\tau_{0}(r)-\tau\right]^{\fr{2}{3+\gamma}},\label{ard}
\eeq
where
\beq
A\equiv 2\pi C (3+\gamma)^{2}, \label{biga}
\eeq
and $\tau_{0}(r)$ is the proper time for complete collapse of a shell
with area radius $R(0,r)=r$, which is given by
\beq
\tau_{0}(r)=\fr{r}{\sqrt{A}}. \label{singularity}
\eeq
Note that $\tau'_{0}>0$, for $r>0$, which is a sufficient condition
for the absence of shell-crossing singularities.

\subsection{Approximate nature of the solution}

A direct computation shows that Eq. (\ref{ard}) satisfies the Einstein equations exactly if $\gamma=0$, or $\gamma=-1$. For other values of $\gamma\in(-1,1)$, the $G_{\theta\theta}$ component of the field equations does not vanish identically. This arises because of the rescaling from $t$ to $\tau$, where an unknown function of $r$ was arbitrarily set to zero, to render $\tau$ and $r$ independent coordinates. This lack of exact solvability of the $G_{\theta\theta}$ component of the field equations is not detrimental, and can be made precise since
\beq
G_{\theta\theta}\propto (1+\gamma)\gamma r^{2(1+\gamma)/(3+\gamma)}F(t,r),
\eeq
where $F(t,0)\propto t^{-(5+3\gamma)/(3+\gamma)}$. Since $\gamma\in(-1,1)$, $G_{\theta\theta}$ is a monotonically increasing function of $r$, which can be made arbitrarily close to zero by considering an arbitrarily small neighborhood of $r=0$, which is precisely the worldline of the central singularity. Therefore, by restricting ourselves to the vicinity of $r=0$, Einstein's equations are satisfied to an arbitrarily small error in the region of interest.

Thus, although not an exact solution for general $\gamma$, our solution satisfies Einstein's equations to arbitrary precision, and, because it has the correct (exact) limiting behavior for $\gamma=-1$ and $\gamma=0$, it provides a reliable test-bed for the study of the role of radial pressure in gravitational collapse. In addition to being an exact solution for $\gamma=-1,0$, Eq. (\ref{ard}) captures all the relevant features of well-posed dust collapse for $\gamma\in(-1,1)$: implosion, satisfaction of the weak energy condition, regular initial metric, and trapped-surface-free initial slice. Another example of a test-bed metric that is not an exact solution of Einstein's equations, but is useful for its physical content, is the Thorne test-bed metric for inspiralling binaries \cite{thorne98}. Such non-exact solutions are quite valuable in giving analytical, if qualitative, insights into physics that otherwise could only be explored numerically.

\section{Conditions for singularity formation and visibility}

The Kretschmann curvature scalar, ${\mathcal K}\equiv R_{abcd}R^{abcd}$, is
\beq
{\mathcal K}=448\pi^{2}C^{2}\fr{1}{R^{4}}\left(\fr{r}{R}\right)^{2(1+\gamma)},
\eeq
which diverges at $R=0$, thereby signaling the existence of a
curvature singularity at a time $\tau_{0}$, where
$R(\tau_{0},r)=0$. Choosing initial data with $\dot{R}(\tau_{\rm i},r)<0$ [note
that the choice $k=0$ precludes the $\tau=0$ slice from being a moment of 
time-symmetry, with $\dot{R}(0,r)=0$], a sufficient condition for collapse is
$\ddot{R}(\tau,r)\leq0$, for $\tau_{\rm i}\leq\tau\leq\tau_{0}$. From
Eq. (\ref{Rdt}), we have
\beq
\ddot{R}(\tau,r)=-\fr{4\pi C}{r}(1+\gamma)\left(\fr{r}{R}\right)^{2+\gamma}.
\eeq
Hence, the condition for complete collapse of a shell $r$ is
\beq
\gamma\geq-1,
\eeq
which is always satisfied. Hence, provided $\dot{R}(\tau_{\rm i}<0,r)<0$, implosion
always occurs, and the formation of a central curvature singularity
is inevitable. Physically, our model corresponds to the initial radial
pressure profile
\beq
\Pi(\tau_{\rm i},r)=\gamma C\left[\fr{r^{1+\gamma}}{R^{3+\gamma}}\right]_{\tau=\tau_{\rm i}}
<\fr{\gamma C}{r^{2}}=\Pi(0,r). \label{radpre}
\eeq
As collapse proceeds, $\Pi$ becomes larger ($\dot{\Pi}>0$), but the
pressure build-up is never enough to halt collapse ($\ddot{R}<0$). To
this extent, this model is particularly useful for examining central
curvature singularities---they always form.

In order to study the collapse of a finite spherical body, we have to introduce a cut-off at some finite coordinate radius $r_{\rm c}$, and match the solution along the timelike three-surface $\Sigma_{\rm c}: \{r=r_{\rm c}\}$ thus defined to a Schwarzschild exterior. Formally, this is achieved by imposing the standard Darmois-Israel junction conditions \cite{darmois-israel} to enforce continuity of the metric and extrinsic curvature along $\Sigma_{\rm c}$. Since our main focus is the central singularity and its properties, we shall not go into the details of the matching and dynamics of the junction surface; rather, we present qualitative arguments to show that such matching can be achieved and that it does not affect the behavior of the solution in the vicinity of the singularity. First, we note that the pressure falls off as $r^{-2}$, and thus we can consider a sufficiently large $r_{\rm c}$, such that $\Pi(\tau,r_{\rm c})\ll1$ (in appropriate mass-type units). If the cut-off at $r_{\rm c}$ is sharp (e.g. step-function), one of the components of the extrinsic curvature will fail to be $C^{0}$ on $\Sigma_{\rm c}$. This can be remedied by considering a thin shell of matter, $r\in(r_{\rm c},r_{\rm c}+\epsilon)$, where the radial pressure (hence the density) decreases smoothly from $r_{\rm c}$ to $r_{\rm c}+\epsilon$, where it vanishes. In this case, the matching with a Schwarzschild exterior is trivial at $r_{\rm c}+\epsilon$, and one only needs to be concerned with the influence of the thin shell on the central singularity. This thin shell can be thought of as a perturbation that travels at the local sound speed $\sqrt{\gamma}$ \cite{ori&piran87&90}. Hence, provided we choose $r_{\rm c}\gtrsim{\mathcal O}(\sqrt{\gamma}\tau_{\rm i})$, any hydrodynamic perturbations will arrive at the center {\em after} the singularity has formed. In addition, from a causal viewpoint, the existence of the thin shell will only introduce a discontinuity on the AH, when the latter crosses the worldline of $r_{\rm c}$---the AH effectively ``jumps out''---collapsing afterwards to the central singularity.

\subsection{Apparent horizon}

The singularity curve is given by Eq. (\ref{singularity}). The
evolution of the apparent horizon can give insight into the causal
structure near the singularity. In the adopted spherical coordinates,
the apparent horizon
(AH), which is the outer boundary of a region containing trapped surfaces, is
given by $2m(\tau_{\rm ah}(r),r)=R(\tau_{\rm ah}(r),r)$. This, from
Eqs. (\ref{mass}) and (\ref{ard}), gives
\beq
\tau_{\rm ah}(r) = \tau_{0}(r)
\left[1-\Theta^{(3+\gamma)/(1+\gamma)}\right], \label{app}
\eeq
with
\beq
\Theta\equiv\sqrt{8 \pi C},
\eeq
where the exponent $(3+\gamma)/(1+\gamma)\in(2,+\infty)$, for
$-1<\gamma<1$.

For non-central shells, the mass function $m(\tau,r)$ is positive definite and
finite on any regular surface and, from Eq. (\ref{mdt0}), increasing
afterwards ($\dot{m}>0$). Therefore, for collapsing configurations,
the quantity $2m/R$ increases monotonically, from its initial value
(less than unity, which  is the condition for non-existence of trapped
surfaces for regular initial data), through unity (whence becoming
trapped),  before diverging positively at the
singular surface, $\tau=\tau_{0}(r)$. The condition for absence of
trapped surfaces on a given initial slice, $2m/R<1$ on any $\tau_{\rm i} < 0$
hypersurface, reads
\beq
\fr{2m}{R}= 8\pi C\left(\fr{r}{R}\right)^{1+\gamma}<1 \label{condition}.
\eeq
Since $(r/R)^{1+\gamma}<1$, $\forall\, \tau_{\rm i} < 0$ and $\gamma > -1$,
the above condition is satisfied provided
\beq
{8\pi C}\leq1, \label{cc}
\eeq
with strict inequality for the $\gamma=-1$ limiting case.
Thus, we have  $\tau_{\rm ah}(r)\leq\tau_{0}$, where the inequality
saturates for $r=0$. Therefore, as expected, non-central shells are strongly
censored and the singularity cannot be visible even locally. Only the
central $r=0$ singularity can be naked. Accordingly, we shall hereafter discuss
the central  ($r=0$) singularity  only.

\section{Visibility}

To show that a singularity is (at least locally) naked, one has to show the
existence of non-spacelike future-directed outgoing
geodesics with their past endpoint at the
singularity. In geometric terms, this is equivalent to requiring that
the area radius increases along such geodesics, i.e., $(dR/dr)_{\rm ORNG}>0$.

The equation for radial null geodesics (RNGs) is
\bqa
\left(\fr{d\tau}{dr}\right)_{\rm RNG}&=&\pm R' \nonumber \\
&=&\pm
\fr{R}{r}\fr{1}{3+\gamma}\left[1+\gamma+2\left(1-\fr{\tau}{\tau_{0}}\right)^{-1}\right],  
\label{rng}
\eqa
where the plus or minus sign corresponds to outgoing or ingoing RNGs,
respectively. Along outgoing RNGs we have
\beq
\fr{dR}{dr}=R'+\dot{R}\left(\fr{d\tau}{dr}\right)_{\rm ORNG}=R'(1+\dot{R}).
\eeq
Using the standard procedure \cite{joshi&dwivedi93}, we introduce the
auxiliary variables $u$, $X$:
\bqa
u&\equiv&r^{\alpha}\;,\;\alpha>0, \label{udef} \\
X&\equiv&\fr{R}{u}. \label{Xdef}
\eqa
In the limit of approach to the singularity we have
\bqa
X_{0}&\equiv&\lim_{R\rightarrow0,u\rightarrow0}\fr{R}{u}=
\lim_{R\rightarrow0,u\rightarrow0}\fr{dR}{du} \nonumber \\
&=&\lim_{R\rightarrow0,r\rightarrow0}\fr{1}{\alpha
r^{\alpha-1}}\fr{dR}{dr} \nonumber \\
&=&\lim_{R\rightarrow0,r\rightarrow0}\fr{1}{\alpha
r^{\alpha-1}}R'(1+\dot{R}).
\eqa
$X_{0}$ is the value of the tangent to the geodesic on the $\{u,X\}$
plane, at the singularity. If $X_{0}\in{\Bbb R}^{+}$, the singularity
is at least locally naked,
and covered otherwise.
Now, from Eq. (\ref{ard}) we have
\beq
\tau(R,r)=\fr{r}{\sqrt{A}}\left[1- \left(\frac{R}{r}\right)^{\fr{3+\gamma}{2}}\right],
\eeq
and
\beq
R'(R,r)=\fr{R}{r(3+\gamma)}\left[1+\gamma+2\left(\fr{r}{R}\right)^{\fr{3+\gamma}{2}}\right].
\eeq
Hence, near the singularity, $\tau_0(0)$, we obtain
\bqa
X_{0}&=&\lim_{r\rightarrow 0}\frac{X}{\alpha(3+\gamma)}
\left[1+\gamma+2\fr{r^{(1-\alpha)(3+\gamma)/2}}{X^{(3+\gamma)/2}}\right]
\times \nonumber \\
&& \left[1-\Theta\frac{r^{(1-\alpha)(1+\gamma)/2}}{X^{(1+\gamma)/2}}\right]
. \label{X0}
\eqa
Clearly, for $X_{0}$ to be finite at $r=0$ we must have $\alpha\leq1$,
since $\gamma\in(-1,1)$. However, if $0<\alpha<1$, $X$ vanishes identically 
on the $\tau=0$ surface; hence, we must have $\alpha=1$, such that $X_{0}$ 
is positive definite on the singular surface.

A self-consistent solution exists if $\alpha=1$. In this case, the
visibility of the singularity depends on the existence of positive
roots of the equation
\beq
X_0^{2+\gamma} + \fr{1+\gamma}{2}\Theta X_0^{(3+\gamma)/2} -X_0^{(1+\gamma)/2}
+\Theta =0. \label{alg}
\eeq
Since $-1<\gamma<1$, this algebraic equation is not a polynomial, and
thus solutions cannot in general be found analytically. We note, however,
the formal similarity with the corresponding equation for the general
spherically symmetric dust case, where only the lowest-order  term appears
with a negative coefficient~\cite{jhingan}.

Even though the existence of real positive roots cannot be determined
for general $\gamma$, one can analytically examine a sufficiently general class of
initial data and draw conclusions from it. Take $\gamma\in[0,1)$ and
consider a class of initial data given by arbitrary $C$ [subject to
the constraint (\ref{cc})] and $\gamma=1/(2n)$, where $n\in{\Bbb
N}^{+}\backslash\{1\}$. The subset of initial data thus constructed is
countable and infinite-dimensional, and hence physically
significant. Equation (\ref{alg}) can then be rewritten as
\bqa
&&Z^{4n+1}+\fr{2n+1}{4n}\Theta Z^{(6n+1)/2}-Z^{(2n+1)/2}+\Theta =0, \\
&&Z\equiv X_{0}^{1/(2n)}.
\eqa
This is a polynomial of odd-degree and hence it has at least one real
root. Now, since $6n+1$ and $2n+1$ are odd, the second and third terms
contain {\em
ostensively} the square-root of $Z$. It then follows that the real
solution(s) must be positive. For this class of initial data there always
exists at least one positive real root of Eq. (\ref{alg}), and the
central singularity is therefore at least locally naked. An analogous
analysis for $\gamma\in(-1,0]$ yields the same results.

A full numerical evaluation of all the real positive roots of
Eq. (\ref{alg}) is shown in Fig. 1, which covers the entire initial
data space, $\{\Theta\in(0,1]\}\otimes\{\gamma\in(-1,1)\}$.

\begin{figure}
\begin{center}
\epsfxsize=15pc
\epsffile{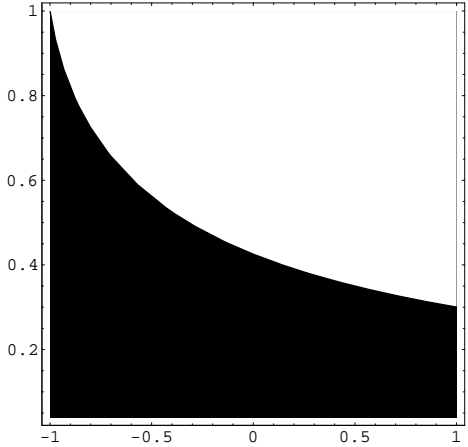}
\end{center}
\caption{Parameter space for the occurrence of naked
singularities. The density increases along the vertical axis ($\Theta$), 
and radial pressure along the horizontal axis ($\gamma$). The shaded 
region corresponds to configurations that collapse to a visible singularity. 
Configurations with initial parameters in the white region end up in black 
holes and contain only spacelike singularities. \label{fig1}}
\end{figure}

It is instructive to examine the following special cases analytically,
as they represent the maximum positive, negative, and zero pressure
cases, respectively.

The limiting case $\gamma=1$ reduces to the cubic polynomial:
\beq
Z^{3}+\Theta Z^{2}-Z+\Theta =0,
\eeq
whose solutions are trivial to analyze, for a given $\Theta$. Standard
methods from polynomial theory show that there is a critical value
$\Theta_{\rm max}=0.300283$, above which there are no positive real
solutions. For $0\leq\Theta<\Theta_{\rm max}$, there are three real roots,
two of which are positive. This case corresponds to the stiffest
equation of state allowed. Even in such a case, visible singularities
occur.

The case $\gamma=-1$ is trivial:
\beq
X_{0}=1-\Theta\geq0.
\eeq
Hence, the singularity is always visible, since in this case $\Theta<1$. 
This is consistent
with the notion that negative radial pressure can lead to
visible singularities that would otherwise be covered \cite{cooperstock}.

The $\gamma = 0$ case corresponds to a subclass of general inhomogeneous dust
models~\cite{PSJ}, and will be dealt with separately in Sec. VI.

There are three relevant regions in the parameter space, defined by
critical values of the strength of the initial density profile,
$\Theta$ (or, equivalently, $C$):

Region I: $\Theta>0.425343$. In this region all the singularities are
covered, for positive values of the radial pressure. Collapse
always ends up in a black hole covering a spacelike singularity, for
$\gamma>0$ and $\Theta>0.425343$. For negative values of the radial
pressure, both black holes and naked singularities can form. In the
$\gamma=-1$ limit, collapse always leads to a visible central
singularity. The more negative the pressure, the larger the number of configurations
ending up in naked singularities.

Region II: $0.425343\leq\Theta\leq0.300283$. For negative pressure,
collapse always ends up in a visible singularity. For positive
pressure, there is a
critical value, $\Theta_{\rm c}$, below which singularities are
visible. When the radial pressure is sufficiently large
($\Theta>\Theta_{\rm c}$), the
singularities are strongly censored and a black hole is the only final
state of collapse. This is in marked contrast with what happens with
tangential stresses, which tend to uncover the singularity.

Region III: $\Theta<0.300283$. Collapse always leads to a visible
singularity, irrespective of the radial pressure. In this case, the
magnitude of the radial pressure is bounded from above, and it is not
enough to censor the singularity. This region shows that radial
pressure alone cannot rule out naked singularities.

\subsection{Existence of an infinite number of ORNGs}

Each real positive root of Eq. (\ref{alg}) uniquely defines an ORNG
with past endpoint at the singularity, and a definite tangent value on
the $\{u,X\}$ plane. To examine the possibility of a {\em family} of
ORNGs emanating from the singularity, we briefly describe the method
of Joshi and Dwivedi \cite{joshi&dwivedi93} and refer the reader to
the original reference for further details.

Let us consider the ORNG equation on the $\{u,X\}$ plane:
\beq
\fr{dX}{du}=\fr{1}{u}\left(\fr{dR}{du}-X\right)=\fr{1}{u}\left[U(X,u)-X\right],
\eeq
where
$$
X_{0}=\lim_{R\rightarrow0,u\rightarrow0} \fr{R}{u}=\lim_{R\rightarrow0,u\rightarrow0} U(X,u).
$$
Now, let us rewrite the algebraic Eq. (\ref{alg}) as
\beq
V(X_{0})\equiv(X_{0}-X^{*}_{0})(f_{0}-1)+f(X_{0})=0,
\eeq
where $X_{0}^{*}$ is one of the real positive roots, $f_{0}$ is a
constant, and the function $f(X_{0})$ contains higher-order terms in
$(X_{0}-X_{0}^{*})$. We have then
\beq
\fr{dX_{0}}{du}=\fr{1}{u}\left[(X_{0}-X_{0}^{*})(f_{0}-1)+S\right], \label{gex}
\eeq
where $S(X,u)\equiv U(X,u)-U(X,0)+h(X_{0})$ vanishes at the
singularity. Using the integrating factor $u^{1-f_{0}}$,
Eq. (\ref{gex}) integrates to
\beq
X_{0}=X_{0}^{*}+Bu^{f_{0}-1}+u^{f_{0}-1}\int Su^{-f_{0}}du,
\eeq
where $B$ is a constant of integration that labels different geodesics.
The last term always vanishes in the limit of approach to the
singularity ($u\rightarrow0$), independently of $f_{0}$. Clearly, if
$f_{0}<1$, the second term diverges, unless $B=0$, which is
required by the existence of the root $X_{0}^{*}$. In this case, a
single ORNG departs from the singularity. However, if $f_{0}>1$, the
second term always vanishes at the singularity, irrespective of $B$,
and thus an infinite number of ORNGs (parameterized by $B$) departs
from the singularity. Now, note that if Eq. (\ref{alg})---and thus
$V(X_{0})$---has (at least) two real positive roots, it follows that
$(dV/dX_{0})=f_{0}-1<0$ along one of the roots, and $f_{0}-1>0$ along the
other. Hence, in such a case, there is a one-parameter family of ORNG
leaving the singularity. A straightforward numerical analysis of the
roots of Eq. (\ref{alg}), shows that for initial data in the shaded
region (excluding its interface boundary; cf. Fig 1.), there are always
two real positive roots (cf. Fig. 2).

\begin{figure}
\begin{center}
\epsfxsize=15pc
\epsfysize=29pc
\epsffile{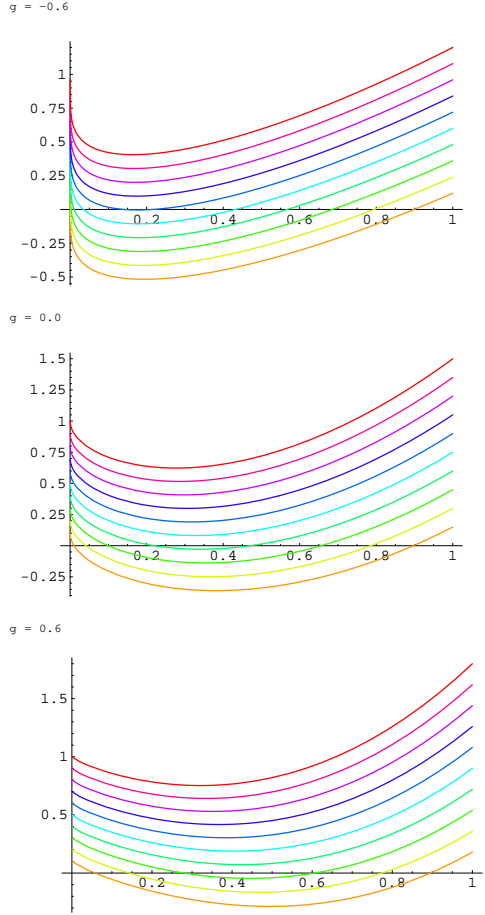}
\end{center}
\caption{Sample plots for roots of the algebraic Eq. (\ref{alg}): from top to bottom, for $\gamma=-0.6, 0, 0.6$, respectively. The numerical value of Eq. (\ref{alg}) is plotted on the vertical axes, and $X_{0}$ along the horizontal axes. For a given graph (i.e. $\gamma$), each curve corresponds to a choice for $\Theta$; one sees that there is a critical value $\Theta_{\rm c}$, below which there are always two real positive roots. \label{fig2}}
\end{figure}

Thus, we conclude that there are infinite number
of ORNGs with past endpoint at the singularity.

\section{Curvature strength}

A defining property of the physical seriousness of a singularity is
its curvature strength. A singularity is said to be gravitationally
strong in the sense of Tipler \cite{tipler77} if every collapsing
volume element is crushed to zero at the singularity, and weak
otherwise (i.e., if it remains finite). It is generally
believed---although not yet proven \cite{clarke93}---that spacetime is
geodesically incomplete at a strong singularity, but extendible
through a weak one \cite{tipler77,tipler&clarke&ellis80}.

A precise characterization of Tipler strong singularities has been
given by Clarke and Kr\' olak \cite{clarke&krolak85}, who proposed
(among other conditions) the {\em limiting strong focusing condition}:
There is at least one non-spacelike geodesic, with tangent vector $K^{a}$ and
affine parameter $\lambda$ (with $\lambda=0$ at the singularity),
along which the scalar $\Psi\equiv R_{ab}K^{a}K^{b}$ satisfies
\beq
\lim_{\lambda\rightarrow0} \lambda^{2}\Psi >0. \label{strong}
\eeq
This is a sufficient condition for the singularity to be Tipler strong
and corresponds to the vanishing of any two-form (for null geodesics),
or three-form (for timelike geodesics) defined along such a geodesic,
at the singularity, due to unbounded curvature growth.

\subsection{Null geodesics}

Let us now consider an ORNG with tangent $K^{a}=(K^{\tau},K^{r},0,0)$,
where $K^{a}=\fr{dx^{a}}{d\lambda}$, and
\beq
K^{\tau}\equiv \fr{F}{R}=R'K^{r}, \label{kt}
\eeq
where $F$ can be written as an explicit function of the affine
parameter, $F=F(\lambda)$, obeying the differential equation (which
follows from the geodesic equation, $K^{a}\nabla_{a}K^{b}=0$):
\beq
\fr{dF}{d\lambda}+F^{2}\left(\fr{\dot{R}'}{RR'}-\fr{\dot{R}}{R^{2}}-\fr{1}{R}\right)=0. \label{dfdl}
\eeq
In our case, we have
\bqa
\Psi&=&\fr{2m'}{R^{2}R'}(K^{\tau})^{2}=\Theta^{2}\left(\fr{r}{R}\right)^{1+\gamma}\fr{(K^{\tau})^{2}}{R^{2}} \nonumber \\
&=&\Theta^{2}X^{-5-\gamma}\fr{F^{2}}{r^{4}},
\eqa
where Eqs. (\ref{udef})-(\ref{Xdef}), and (\ref{dfdl}) were used. Hence,
\bqa
\lim_{\lambda\rightarrow0} \lambda^{2}\Psi &=& \Theta^{2}X_{0}^{-5-\gamma} \lim_{\lambda\rightarrow0} \fr{\lambda^{2}}{F^{-2}r^{4}} \nonumber \\
&=& \Theta^{2}X_{0}^{-5-\gamma} \lim_{\lambda\rightarrow0} \fr{1}{\Omega^{2} r^{4}} \nonumber \\
&=& X_{0}^{2},
\eqa
where l'H\^ opital's rule was used twice, and [from Eq. (\ref{dfdl})]
\beq
\Omega\equiv\fr{1}{Xr^{2}}\left(\Theta X^{-(5+\gamma)/2}-r\right).
\eeq
Since $X_{0}>0$ always exists (cf. previous subsection), we conclude
that the singularity is Tipler strong {\em independently} of the
details of the initial data. We note, also, that there are infinite
number of ORNGs with past endpoint at the singularity, each one of
them defined by a unique tangent vector (on the $\{u,R\}$ plane)
$X_{0}$, which is determined from the initial data ($C$ and
$\gamma$). This constitutes a rather robust result, in that (i) the
singularity is Tipler strong irrespective of the initial data, and
(ii) it is so along an infinite number of ORNGs terminating at the
singularity in the past.

\subsection{Timelike geodesics}

Let us consider radial timelike geodesics (RTGs), with tangent vector
$\xi^{a}=\fr{dx^{a}}{d\lambda}$, where $\lambda$ is proper time along
the geodesic, and
\bqa
&&\xi^{\tau}=\pm\sqrt{1+R'^{2}(\xi^{r})^{2}}, \\
&&\dot{\xi^{r}}R'+2\xi^{r}\dot{R}'+\fr{\xi^{r}}{\xi^{\tau}} 
(\xi^{r})'R'+\fr{(\xi^{r})^{2}}{\xi^{\tau}}R''=0, \label{ge}
\eqa
where the first equation is simply $\xi^{a}\xi_{a}=-1$, and the second
follows from the geodesic equation. By inspection, one sees that
Eq. (\ref{ge}) admits the trivial solution
\bqa
\xi^{\tau}&=&\pm1, \label{xitau} \\
\xi^{r}&=&0, \label{xir}
\eqa
which leads to
\bqa
\tau=\tau_{\rm i}\pm(\lambda-\lambda_{0}), \label{timet} \\
r=r_{0}=\mbox{const.},
\eqa
where the plus or minus sign refers to outgoing or ingoing RTGs,
respectively. The outgoing RTG departing from the singularity is given
by $r=0$, and $\tau=\tau_{0}+\lambda-\lambda_{0}$, and thus does not
belong in the spacetime.

The ingoing RTG is given by $r=0$,
$\tau=\tau_{0}-\lambda+\lambda_{0}$, where $\tau_{\rm i}=\tau_{\rm
0}(0)=0$ is the time at which the RTG arrives at the singularity.  For
this RTG, we have
\beq
\Psi\equiv R_{ab}\xi^{a}\xi^{b}=R_{\tau\tau}(\xi^{\tau})^{2}=\fr{m'}{R^{2}R'}=4\pi C\fr{r^{1+\gamma}}{R^{3+\gamma}}.
\eeq
From Eqs. (\ref{ard}) and (\ref{timet}) we get
\beq
R=\left[Ar^{1+\gamma}(\lambda-\lambda_{0})^{2}\right]^{\fr{1}{3+\gamma}},
\eeq
which gives
\beq
\Psi=\fr{2}{(\lambda-\lambda_{0})^{2}(3+\gamma)^{2}}.
\eeq
Thus
\beq
\lim_{\lambda\rightarrow\lambda_{0}} (\lambda-\lambda_{0})^{2}\Psi=\fr{2}{(3+\gamma)^{2}}.
\eeq
Hence, the singularity is always Tipler strong along the RTG, irrespective of the initial data.

\section{The dust limit}

When $\gamma=0$, we have, from Eqs. (\ref{ard}) and (\ref{mass}):
\beq
R(\tau,r)=\left(\fr{9m}{2}\right)^{\fr{1}{3}}(\tau_{0}-\tau)^{\fr{2}{3}},
\label{RLTB}
\eeq
where
\beq
m=m(r)=4\pi Cr=4\pi C\rho(0,r) r^{3}. \label{mass0}
\eeq
Equation (\ref{RLTB}) is exactly the same as the one in the marginally
bound inhomogeneous dust case \cite{PSJ}. However, as can be seen from
Eq. (\ref{mass0}),
the $\gamma=0$ limit of our model is a particular case of the
general inhomogeneous case. The initial mass function cannot be arbitrarily
specified, with the only free parameter being the constant $C$.

At $\tau=\tau_{0}(r)$, a curvature singularity forms. Events with $r>0$
are spacelike and thus covered (cf. Sec. III). Of potential interest
is the $r=0$
singularity. The visibility of such a singularity is determined by the
existence of real positive roots of the algebraic equation
(\ref{alg}). In this case, this equation reduces to the quartic:
\beq
P(Y)=Y^{4}+\fr{\Theta}{2}Y^{3}-Y+\Theta=0, \label{quart}
\eeq
where $Y\equiv \sqrt{X_{0}}$. Standard results from polynomial theory
can be used to show that there is a maximum value $\Theta_{\rm
max}=0.425343$, above which all the roots of the quartic are
complex. This result agrees with that of Joshi and Singh (JS) for pure dust
\cite{PSJ}, upon identification of the mass function in the two
approaches, whence ones finds [cf. Eqs. (13)-(15) in Ref. \cite{PSJ}]:
\beq
\Theta_{\rm max}^{2} = 8\pi C = \lambda_{\rm JS} = 16 \beta_{\rm JS}=0.180917.
\eeq

For $\Theta\in(0,\Theta_{\rm max}]$, there are two complex
conjugate roots and two real positive-definite roots. (When
$\Theta=0$, we obtain the trivial solution $Y=X_{0}=0$ and $Y=X_{0}=1$,
where the former is a degenerate case, where the tangent
to the ORNG vanishes on the $\{u,X\}$ plane).

As an alternative way of showing the result above,
note that Eq. (\ref{quart}) defines a two-dimensional surface that
intersects the $P=0$ plane along a curve $\Theta_{0}(Y)$, given by
solving Eq. (\ref{quart}) for $\Theta$:
$$
\Theta_{0}(Y)=-2\fr{Y(Y^{3}-1)}{Y^{3}+2}.
$$
On the $P=0$ plane, $\Theta_{0}$ has an absolute maximum at
$Y_{*}=0.581029$, with
$\Theta_{0}(Y_{*})=0.425343$.

From the results of Sec. V, we find that the central singularities in
the pure dust limit are Tipler strong along an infinite number of
ORNGs and at least one IRTG, irrespective of the initial data, in
agreement with earlier results for general inhomogeneous dust
\cite{deshingkar&joshi&dwivedi99}.

\section{Discussion and Conclusions}

The spherical symmetric collapse of dust with non-vanishing radial
pressure $\Pi$, and vanishing tangential stresses was studied. When
restricted to marginally bound configurations ($k=0$), the field
equations were analytically integrated in closed form. The $k=0$
ansatz, together with the equation of state $\Pi=\gamma\rho$, implies
that the initial data consists of two free parameters, $\Theta$ and
$\gamma$, which measure the ``strength'' of the initial density and
pressure profiles, respectively.

As in the pressureless dust case, a central singularity forms, which
is Tipler strong along an infinite number of ORNGs, and at least one
IRTG. Even though, in the timelike case, curvature growth was studied
along a particular geodesic, using similar causal structure arguments
to those of Deshingkar, Joshi, and Dwivedi
\cite{deshingkar&joshi&dwivedi99}, we can qualitatively show the
existence of an infinite number of IRTG with future endpoint at the
singularity, as follows. The existence of a future-directed timelike geodesic
$\sigma(\lambda)$ with future endpoint at the singularity (proved in
Sec. V), implies that its chronological past $I^{-}[\sigma]$ is a {\em
timelike indecomposable past set} (TIP) \cite{gerochetal72}. In other
words, the TIP is generated by $\sigma$, which is timelike and
future-inextendible. The locally naked singularity itself---which is
the future endpoint of $\sigma$, $\sigma(\lambda_{0})$---constitutes a
singular TIP which contains the past $I^{-}(p)$ of any point
$p\in\sigma\backslash\{\sigma(\tau_{0})\}$ \cite{penrose78}. Consider
now another point $q\not\in\sigma$ in the past of the singularity and
in the chronological past of $p$. Since $I^{-}(q)\subset I^{-}(p)$, it
follows that there exists a timelike curve from $p$ to $q$, say
$\zeta$, satisfying $I^{-}[\zeta]\subset I^{-}[\sigma]$. Consider then
a small compact ball ${\mathcal B}$---with a suitably defined
``radius'', e.g., proper geodesic distance along $\sigma$,
$d_{\sigma}(\lambda,\lambda_{0})$---in the neighborhood of
$\sigma(\tau_{0})$, partially contained in TIP $I^{-}[\sigma]$ and in
the chronological future of $p$, i.e., ${\mathcal B}\cap
I^{-}[\sigma]={\mathcal C}\nsubseteq\emptyset$ and $I^{+}[{\mathcal
C}]\subset I^{+}(p)$. For each point $x\in{\mathcal C}$, there is a
timelike geodesic from $p$ to $x$. Since ${\mathcal C}$ is compact,
there are infinite number of such points that can be joined by
timelike geodesics from $p\in\sigma$. Now, let $p$ be an arbitrary
point on $\sigma$ and take the limit of approach to the singularity,
$d_{\sigma}(\lambda,\lambda_{0})\rightarrow0^{+}$; it then follows
that there are infinite number of future-inextendible timelike
geodesics with future endpoint at the singularity.

More important than the directional behavior of curvature strength, is
the visibility of the singularity. We found an infinite number of
ORNGs with past endpoint at the singularity, which is therefore at
least locally naked. The existence of such ORNGs is independent of the
initial data. This differs from the pressureless innhomogeneous dust
case, where the visibility of the central singularity depends
crucially on differentiability properties of the central energy
density distribution, and it is because our model reduces to a
pressureless dust class with a particular mass function ($m\propto
r$). In our model, the initial energy density distribution has a
single degree of freedom ($\Theta$), and is $C^{\infty}$ everywhere
except at the center, where it diverges as $r^{-2}$. However, the mass
function is everywhere finite (it vanishes at $r=0$), and the metric
is everywhere regular. We remark that the divergence of the energy density at $r=0$ does {\em not} preclude our model from being a useful test-bed for cosmic censorship. The issue of cosmic censorship is the issue of the development of singularities---signalled by the formation of trapped surfaces \cite{hawking&ellis73}---from an initial spacelike slice that does not admit trapped surfaces and on which the induced metric is regular, in the {\em causal future} of that slice. In our model, the initial data ($\Theta$ and $\gamma$) are such that the metric is {\em everywhere regular}, and there are no trapped surfaces on the initial slice. We have shown that, from a large subset of initial data, trapped surfaces---and, consequently, singularities---inevitably form in the causal future of such regular initial slices. Accordingly, whereas this model may not provide a realistic
approximation for the central density, it constitutes a well-defined
test-bed spacetime to examine the isolated effects of radial pressure on the formation and visibility of the central singularity.

We find that positive radial pressure tends to supress visible
singularities, and negative radial pressure tends to uncover them. Our
negative pressure data space complements the results of Cooperstock
{\em et al.} for non-central singularities \cite{cooperstock}, to
include the $r=0$ case.  The most significant result is that,
irrespective of the strength of the radial pressure, there is always a
set of non-zero measure in the parameter space that leads to visible
singularities. Hence, we find that radial pressure alone cannot
eliminate the naked singularity spectrum. We conjecture that this
behavior is not particular to our (marginally bound) model, and that
it will exist in {\em any} spherical system with an equation of state
of the form $\Pi=\gamma\rho$. In addition, since tangential stresses alone tend to
uncover the singularity spectrum, and radial pressure by itself fails
to fully cover it, we further speculate that when the former and
latter are taken together into account, visible singularities will
still persist.

In order to confidently establish the role of radial pressure and the
combined roles of tangential and radial stresses in the final state of
gravitational collapse, further analyses need to be undertaken, of
non-marginally-bound systems, and explicitly including tangential
stresses. Efforts in this direction are currently underway \cite{SG-SJ}.

\section*{Acknowledgments}
The authors are grateful to P. Brady, T. Harada, H. Iguchi, P. Joshi,
H. Kodama, and K. Thorne for helpful discussions and/or
comments. SJ thanks T. Tanaka for help with producing Mathematica figures.
SMCVG acknowledges the support of FCT (Portugal) Grant
PRAXIS XXI-BPD-16301-98, and NSF Grant AST-9731698. SJ acknowledges
the support by Grant-in-Aid for JSPS fellows (No. 00273).

\end{document}